\documentstyle[12pt]{article}
\newcommand{\sslash}{\not \!\! s}
\newcommand{\kslash}{\not \! k}
\newcommand{\pslash}{\not \! p}
\newcommand{\delslash}{\not \! \partial}
\begin{document}

\begin{flushright}{UT-967\\ September 2001}
\end{flushright}
\vskip 0.5 truecm

\begin{center}
{\Large{\bf Lattice Chiral Symmetry and the
Wess-Zumino Model }}
\end{center}
\vskip .5 truecm
\centerline{\bf Kazuo Fujikawa and Masato Ishibashi}
\vskip .4 truecm
\centerline {\it Department of Physics,University of Tokyo}
\centerline {\it Bunkyo-ku,Tokyo 113,Japan}
\vskip 0.5 truecm

\makeatletter
\@addtoreset{equation}{section}
\def\theequation{\thesection.\arabic{equation}}
\makeatother

\begin{abstract}
A lattice regularization of the supersymmetric Wess-Zumino 
model is studied by using Ginsparg-Wilson operators. 
We recognize a certain conflict between the lattice chiral
symmetry and the Majorana condition for  Yukawa 
couplings, or 
in Weyl representation a conflict between the lattice chiral
symmetry and Yukawa couplings.  This conflict is also related, 
though not directly, 
to the fact that the kinetic (K\"{a}hler) term and the 
superpotential
term are clearly distinguished in the continuum Wess-Zumino 
model, whereas these two terms are mixed in the Ginsparg-Wilson 
operators. We illustrate a case where lattice chiral symmetry 
together with naive Bose-Fermi symmetry is imposed by 
preserving a SUSY-like symmetry in the free part of the 
Lagrangian; one-loop level non-renormalization of the 
superpotential is then maintained for finite lattice spacing, 
though the finite parts of  wave function renormalization 
deviate from the supersymmetric value. 
All these properties hold for the general Ginsparg-Wilson 
algebra independently of the detailed construction of lattice
Dirac operators.
\end{abstract}


\section{Introduction}
Supersymmetry is a result of the pursuit of maximal
symmetry between bosons and fermions, and it exhibits 
remarkable dynamical properties\cite{wess-bagger}. A study of
the simplest model due to Wess and Zumino\cite{wess} led to
 the so-called non-renormalization theorem both in the 
 component approach\cite{iliopoulos} and in 
superfield formulation\cite{fujikawa}. As was emphasized in 
the first paper in \cite{fujikawa}, the potential part of the
model is free of any (even finite) renormalization up to
all orders in perturbation when renormalized at 
{\em vanishing momenta}. However, it may be fair to say that a 
satisfactory regularization of general 
supersymmetry, in particular  in component formulation, is still 
missing.

On the other hand, we recently witnessed a remarkable 
progress in lattice gauge theory. A basic algebraic relation 
which lattice Dirac operators should satisfy was clearly 
stated by Ginsparg and Wilson\cite{ginsparg}, and an explicit  
solution to the algebra 
was given\cite{neuberger}. This
solution exhibits quite interesting chiral properties
\cite{hasenfratz}\cite{luscher}\cite{Hasenfratz:1998jp}
 including locality 
properties\cite{hernandez}\cite{neuberger2}\cite{horvath};
the lattice Dirac operator is no more ultra-local but
exponentially local. See, for example, 
Ref.\cite{niedermayer} for  reviews of this development.

It is thus interesting to see what we  learn about the 
regularization of supersymmetry  if one uses the Ginsparg-Wilson
algebra. In this paper, we study this problem by using the 
simplest Wess-Zumino model. There already exist several
papers on this subject\cite{kikukawa}\cite{bietenholz},
but our approach 
differs from those studies in several essential aspects.
Our interest is in the implications of lattice chiral symmetry
satisfied by the Ginsparg-Wilson operators, and we study the 
divergence cancellation by perturbative calculations, setting 
aside the basic issue of supersymmetry 
algebra\cite{kikukawa}\cite{bietenholz}.
The perturbative lattice analysis of Yukawa couplings
is expected to be important when one applies a non-perturbative 
analysis to the gauge theoretical sector of a supersymmetric 
model of elementary particles, just to preserve the consistency 
of various sectors of a single model.

We show that there exists a conflict between the lattice chiral
symmetry and the Majorana condition for  Yukawa couplings, or 
in Weyl representation a conflict between the lattice chiral
symmetry and Yukawa couplings.  
This conflict is also related, though not directly,
to the fact that the kinetic (K\"{a}hler) term and the 
superpotential term are clearly distinguished in the continuum 
Wess-Zumino model 
(in a symbolic superfield notation\cite{fujikawa})
\begin{equation}
S=\int \phi^{\dagger}\phi+\int(m\phi^{2}+g\phi^{3}) + h.c.,
\end{equation}
whereas these two terms are mixed in the Ginsparg-Wilson 
operators, as is explained later. We then illustrate in some
detail a case where lattice chiral symmetry 
together with naive Bose-Fermi symmetry is imposed by sacrificing
the precise Majorana condition but by
preserving a SUSY-like symmetry in the free part of the 
Lagrangian. One-loop level non-renormalization of the 
superpotential is shown to be maintained for finite lattice 
spacing, though the finite parts of  wave function 
renormalization deviate from the supersymmetric value. 
 
A pioneering work of the lattice regularization of 
supersymmetry was initiated by Dondi and Nicolai\cite{dondi}.
Among the past works, our analysis is closely related to the 
work of Bartels and Kramer\cite{bartels}. 
See also the early works on the Wess-Zumino model\cite{elitzur} 
and supersymmetric gauge theory\cite{banks}.
As for the recent analyses of lattice regularization of 
supersymmetric gauge theory on the basis of the Ginsparg-Wilson 
operator (or the domain-wall fermion), see papers in\cite{maru}.
\\
\\
{\bf Brief Summary of the Ginsparg-Wilson Operators}
\\

We are interested in the implications of the 
Ginsparg-Wilson algebra {\em per se} independently of the 
detailed 
properties of lattice Dirac operators. To be more
precise, we analyze the implications of the general
algebraic relation\cite{fujikawa2} 
\begin{equation}
\gamma_{5}(\gamma_{5}D)+(\gamma_{5}D)\gamma_{5}=
2a^{2k+1}(\gamma_{5}D)^{2k+2}
\end{equation}
where $D$ is the lattice Dirac operator and the parameter $a$ 
is the lattice spacing; $k$ stands for 
non-negative integers, and $k=0$ 
corresponds to the conventional Ginsparg-Wilson relation. 
When one defines a hermitian operator $H$ by
\begin{equation}
H=a\gamma_{5}D=H^{\dagger}=aD^{\dagger}\gamma_{5}
\end{equation}
the above algebraic relation is written as
\begin{equation}
\gamma_{5}H+H\gamma_{5}=2H^{2k+2}.
\end{equation}
We can also show
\begin{equation}
\gamma_{5}H^{2}=(\gamma_{5}H+H\gamma_{5})H-
H(\gamma_{5}H+H\gamma_{5})+ H^{2}\gamma_{5}=H^{2}\gamma_{5}
\end{equation}
which implies
\begin{equation}
H^{2}=a^{2}D^{\dagger}D=\gamma_{5}H^{2}\gamma_{5}
=a^{2}DD^{\dagger}.
\end{equation}

When we define
\begin{eqnarray}
&&\Gamma_{5}\equiv \gamma_{5}-H^{2k+1},\nonumber\\
&&\hat{\gamma}_{5}\equiv \gamma_{5}-2H^{2k+1},
\end{eqnarray}
the defining algebra (1.2) is written as
\begin{equation}
\Gamma_{5}H+H\Gamma_{5}=0
\end{equation}
and 
\begin{equation}
(\hat{\gamma}_{5})^{2}=1.
\end{equation}
We can also show the relation
\begin{equation}
\gamma_{5}\Gamma_{5}+\Gamma_{5}\gamma_{5}=2\Gamma_{5}^{2}
=2(1-H^{4k+2})
\end{equation}
which implies $H^{2}\leq 1$.
We next note\cite{niedermayer}   
\begin{equation}
D = P_{+}D\hat{P}_{-} +  P_{-} D\hat{P}_{+}    
\end{equation}
or
\begin{equation}
H= P_{+}H\hat{P}_{-}+ P_{-}H\hat{P}_{+} 
\end{equation}
which implies
\begin{eqnarray}
&&P_{+}H=P_{+}H\hat{P}_{-}=H\hat{P}_{-},\nonumber\\
&&P_{-}H=P_{-}H\hat{P}_{+}=H\hat{P}_{+}.
\end{eqnarray}
Here we defined two projection operators
\begin{eqnarray}
P_{\pm} &=& \frac{1}{2}( 1 \pm \gamma_{5}),\nonumber\\
\hat{P}_{\pm} &=& \frac{1}{2}( 1 \pm \hat{\gamma}_{5})
\end{eqnarray}
which satisfy the relations
\begin{eqnarray}
&&P_{+}\hat{P}_{+}=P_{+}\Gamma_{5},\nonumber\\
&&P_{-}\hat{P}_{-}=-P_{-}\Gamma_{5}.
\end{eqnarray}
The following relations are often used in our calculation
\begin{eqnarray}
&&P_{\pm}\Gamma_{5}H=-H\Gamma_{5}P_{\mp}=\Gamma_{5}HP_{\mp},
\nonumber\\
&&P_{\pm}\Gamma_{5}HP_{\pm}=0
\end{eqnarray}
which follow from $P_{+}H-HP_{-}=H^{2k+2}$ and $P_{+}+P_{-}=1$.

We then define the chiral components\cite{niedermayer}
\begin{equation}
\bar{\psi}_{L,R}= \bar{\psi}P_{\pm}, \ \ 
\psi_{R,L}= \hat{P}_{\pm}\psi
\end{equation}
and the scalar and pseudscalar densities by\cite{chandrasekharan} 
\begin{eqnarray}
S(x) &=& \bar{\psi}_{L}\psi_{R} + \bar{\psi}_{R}\psi_{L}
=\bar{\psi}\gamma_{5}\Gamma_{5}\psi,
\nonumber\\
P(x) &=& \bar{\psi}_{L}\psi_{R} - \bar{\psi}_{R}\psi_{L}
=\bar{\psi}\Gamma_{5}\psi.
\end{eqnarray}

An explicit form of $H$ in momentum representation, which
is local and free of species doublers, is given by\cite{fujikawa3}
\cite{chiu}
\begin{eqnarray}
H(ap_{\mu})&=&\gamma_{5}(\frac{1}{2})^{\frac{k+1}{2k+1}}
(\frac{1}{\sqrt{H^{2}_{W}}})^{\frac{k+1}{2k+1}}
\{(\sqrt{H^{2}_{W}}+M_{k})^{\frac{k+1}{2k+1}}
-(\sqrt{H^{2}_{W}}-M_{k})^{\frac{k}{2k+1}}
\frac{\sslash}{a} \}\nonumber\\
&=&\gamma_{5}(\frac{1}{2})^{\frac{k+1}{2k+1}}
(\frac{1}{\sqrt{F_{(k)}}})^{\frac{k+1}{2k+1}}
\{(\sqrt{F_{(k)}}+\tilde{M}_{k})^{\frac{k+1}{2k+1}}
-(\sqrt{F_{(k)}}-\tilde{M}_{k})^{\frac{k}{2k+1}}
\sslash \}\nonumber\\
&&
\end{eqnarray}
where
\begin{eqnarray}
F_{(k)}&=&(s^{2})^{2k+1}+\tilde{M}_{k}^{2},\nonumber\\
\tilde{M}_{k}&=&[\sum_{\mu}(1-c_{\mu})]^{2k+1}
-m_{0}^{2k+1}
\end{eqnarray}
and
\begin{eqnarray}
&&s_{\mu}=\sin ap_{\mu}\nonumber\\
&&c_{\mu}=\cos ap_{\mu}\nonumber\\
&&\sslash=\gamma^{\mu}\sin ap_{\mu}.
\end{eqnarray}
For $k=0$, this operator is reduced to  Neuberger's overlap 
operator\cite{neuberger}. 
Our Euclidean Dirac matrices are anti-hermitian, 
$(\gamma^{\mu})^{\dagger}=-\gamma^{\mu}$, but the inner 
product is defined to be $s^{2}\geq 0$. Note that $H^{2}$
(and consequently $\Gamma_{5}^{2}$) is independent of Dirac 
matrices
\footnote{The basic algebra implies 
$[\gamma_{5},H^{2}]=0$
and thus $H^{2}$ contains an even number of Dirac matrices. We 
thus expect that free $H^{2}$ is generally independent of Dirac 
matrices 
since it depends on a single momentum and parity conserving.}.
The parameter $m_{0}$ is constrained by $0<m_{0}<2$, and 
$2m^{2k+1}_{0}=1$ gives a proper normalization
of $H$, namely, for an infinitesimal $p_{\mu}$, i.e.,
for $|ap_{\mu}|\ll 1$,
\begin{equation}
H\simeq-\gamma_{5}a\pslash(1+O(ap)^{2})
+\gamma_{5}(\gamma_{5}a\pslash)^{2k+2}
\end{equation}
to be consistent with $H=\gamma_{5}aD$; the last term in the 
rigth-hand side is the 
leading term of chiral symmetry breaking terms.

Most of our calculations in the following are independent of the 
detailed construction of $H$.

\section{Lattice Chiral Symmetry and the Majorana Condition}

The continuum Wess-Zumino model is defined by\cite{iliopoulos}
\begin{eqnarray}
{\cal L}&=&\frac{1}{2}\chi^{T}Ci\delslash\chi 
+ \frac{1}{2}m\chi^{T}C\chi
+g\chi^{T}C(P_{+}\phi P_{+}
+P_{-}\phi^{\dagger} P_{-})\chi\\
&&+\phi^{\dagger}\partial_{\mu}\partial^{\mu}\phi+F^{\dagger}F
+m[F\phi+(F\phi)^{\dagger}]
+g[F\phi^{2}+(F\phi^{2})^{\dagger}]\nonumber  
\end{eqnarray}
where $\chi$ is  a Euclidean Majorana fermion and
\begin{equation}
\phi=(A+iB)/\sqrt{2}.
\end{equation} 
The variable $F$ stands for a complex auxiliary field.
There exists
no Euclidean Majorana fermion in a strict sense, and we follow 
the conventional treatment of a Euclidean Majorana 
fermion\cite{nicolai}. See Ref.\cite{van nieuwenhuizen} for a 
nice  account of Euclidean Majorana fermions.
It is easier to handle a lattice Dirac fermion than a lattice 
Majorana fermion. We thus start with
a lattice Dirac fermion with Yukawa couplings, and later we
discuss the reduction of the Dirac fermion  to a Majorana fermion.

\subsection{Dirac fermions and Yukawa couplings}

We can think of 3 different Lagrangians for the Dirac fermions 
with Yukawa couplings. The first one is the most natural 
one consistent with lattice chiral symmetry, which is softly
broken by the mass term,
\begin{eqnarray}
{\cal L}_{(1)}&=&\bar{\psi}D\psi 
+ m\bar{\psi}\gamma_{5}\Gamma_{5}\psi
+ 2g\bar{\psi}(P_{+}\phi\hat{P}_{+}
+P_{-}\phi^{\dagger}\hat{P}_{-})\psi\nonumber\\
&=&\bar{\psi}_{R}D\psi_{R}+\bar{\psi}_{L}D\psi_{L} 
+ m[\bar{\psi}_{R}\psi_{L}+\bar{\psi}_{L}\psi_{R}]\nonumber\\
&&+ 2g[\bar{\psi}_{R}\phi\psi_{L}
+\bar{\psi}_{L}\phi^{\dagger}\psi_{R}].  
\end{eqnarray}
We fixed the mass term in such a way that it is 
generated by a shift 
$\phi\rightarrow\phi+m/(2g)$ in 
$\phi=(A+iB)/\sqrt{2}$ in the interaction terms; we adopt this 
procedure in the following.
The fermion mass term is then defined by the scalar density 
formed of a fermion bi-linear (1.18). 
The fermion propagator is given by
\begin{equation}
\langle\psi(y)\bar{\psi}(x)\rangle=(-)\frac{a}{H+am\Gamma_{5}}
\gamma_{5}
=(-)\frac{a(H+am\Gamma_{5})}{H^{2}+(am\Gamma_{5})^{2}}\gamma_{5}.
\end{equation} 
Note that $H^{2}$ and $\Gamma_{5}^{2}$ are proportional to a 
$4\times 4$ unit Dirac matrix.

The second possible Lagrangian incorporates the 
continuum chiral symmetry for  Yukawa couplings, which is 
softly broken by the mass term,
\begin{eqnarray}
{\cal L}_{(2)}&=&\bar{\psi}D\psi 
+ m\bar{\psi}\psi
+ 2g\bar{\psi}(P_{+}\phi P_{+}
+P_{-}\phi^{\dagger}P_{-})\psi  
\end{eqnarray}  
but no explicit lattice chiral symmetry even for $m=0$. 
The propagator is given by
\begin{eqnarray}
\langle\psi(y)\bar{\psi}(x)\rangle=(-)\frac{a}{H+am\gamma_{5}}
\gamma_{5}&=&(-)\frac{a(H+am\gamma_{5})}
{H^{2}+am(H\gamma_{5}+\gamma_{5}H)+(am)^{2}}\gamma_{5}\nonumber\\
&=&(-)\frac{a(H+am\gamma_{5})}
{H^{2}+2amH^{2k+2}+(am)^{2}}\gamma_{5}
\end{eqnarray}
where we used the defining algebra of the Ginsparg-Wilson 
operator\footnote{This propagator approaches 
for large $k$ to $\sim(-)\frac{a(H+am\gamma_{5})}
{H^{2}+(am)^{2}}\gamma_{5}$,
since $H^{2}\leq 1$ and almost all the eigenvalues of $H$ are 
less than unity. }. This propagator has a pole structure 
different from the one naively expected on the basis of the 
Lagrangian (2.5).

Yet another Lagrangian, which is suggested by the analysis of
lattice chiral gauge theory with a real representation of 
gauge group\cite{suzuki}, is
\begin{eqnarray}
{\cal L}_{(3)}&=&\bar{\psi}D\psi 
+ im\bar{\psi}\gamma_{5}\psi
+2ig\bar{\psi}\gamma_{5}(P_{+}\phi P_{+}
+P_{-}\phi^{\dagger}P_{-})\psi.
\end{eqnarray}
Note that we fix the mass term in such a way that it is 
generated by a shift $\phi\rightarrow\phi+m/(2g)$ in 
the interaction terms. (To be precise, the analysis in 
\cite{suzuki}
fixes the mass term, and we fixed the interaction terms 
accordingly.) 
This Lagrangian incorporates a certain aspect of lattice 
chiral symmetry as is seen by evaluating the propagator
\begin{equation}
\langle\psi(y)\bar{\psi}(x)\rangle=(-)\frac{a}{H+iam}
\gamma_{5}=(-)\frac{a(H-iam)}{H^{2}+(am)^{2}}\gamma_{5}
\end{equation} 
which has a pole structure naively expected on the basis of 
the Lagrangian. 
The factor $i\gamma_{5}$ in the Lagrangian is eliminated by a 
$\pi/4$ chiral rotation up to chiral anomaly, which is absent in 
the present model,
\begin{eqnarray}
&&\psi\rightarrow e^{-i(\pi/4)\gamma_{5}}\psi,\nonumber\\
&&\bar{\psi}\rightarrow \bar{\psi}e^{-i(\pi/4)\gamma_{5}}
\end{eqnarray}
if the kinetic term is invariant under the continuum chiral 
symmetry. However, the Ginsparg-Wilson operator is not invariant 
under the continuun chiral symmetry, and thus the factor 
$\gamma_{5}$ has an intrinsic meaning. In fact we later 
show that this scheme violates $CP$ symmetry.

\subsection{Majorana reduction and supersymmetry}

We now define a Majorana fermion starting with the Lagrangian 
for a Dirac fermion. 
In the analysis of a Majorana fermion, the transpose
operation is essential. 
We employ the convention
\begin{eqnarray}
&&C\gamma^{\mu}C^{-1}=-(\gamma^{\mu})^{T},\\ 
&&C\gamma_{5}C^{-1}=\gamma^{T}_{5},\\ 
&&C^{\dagger}C=1,\ \  C^{T}=-C.
\end{eqnarray}
We then have
\begin{eqnarray}
&&C^{T}=-C\nonumber\\
&&(C\gamma_{5})^{T}=-C\gamma_{5}\nonumber\\
&&(CD)^{T}=-CD, \ \ \ (C\gamma_{5}\Gamma_{5})^{T}
=-C\gamma_{5}\Gamma_{5}\\
&&(CH)^{T}=(C\gamma_{5}aD)^{T}=-CaD\gamma_{5}\neq 
-CH, \ \ \ (C\Gamma_{5})^{T}\neq -C\Gamma_{5}.\nonumber
\end{eqnarray}
The basic idea of the Majorana reduction is to decompose a
Dirac fermion $\psi$ into a sum of two Majorana fermions
$\chi$ and $\eta$, which are Grassmann numbers, as
\begin{eqnarray}
&&\psi=(\chi+i\eta)/\sqrt{2}\nonumber\\
&&\bar{\psi}=(\chi^{T}C-i\eta^{T}C)/\sqrt{2}.
\end{eqnarray}

\subsubsection{Majorana fermion with lattice chiral symmetry}

We start with the Dirac fermion with lattice chiral symmetry 
(2.3)
and apply the above substitution (2.14). We naively expect
\begin{eqnarray}
{\cal L}_{(1)}&=&\frac{1}{2}\chi^{T}CD\chi 
+ \frac{1}{2}m\chi^{T}C\gamma_{5}\Gamma_{5}\chi
+ g\chi^{T}C(P_{+}\phi\hat{P}_{+}
+P_{-}\phi^{\dagger}\hat{P}_{-})\chi\nonumber\\
&+&\frac{1}{2}\eta^{T}CD\eta 
+ \frac{1}{2}m\eta^{T}C\gamma_{5}\Gamma_{5}\eta
+ g\eta^{T}C(P_{+}\phi\hat{P}_{+}
+P_{-}\phi^{\dagger}\hat{P}_{-})\eta  
\end{eqnarray}
but this actually fails\footnote{If $(CO)^{T}=-CO$ for a general
operator $O$, the cross
term vanishes $\eta^{T}CO\chi-\chi^{T}CO\eta=0$ by using the 
anti-commuting property of $\chi$ and $\eta$.} if one recalls
\begin{equation}
(P_{+}\phi\hat{P}_{+}
+P_{-}\phi^{\dagger}\hat{P}_{-})=\frac{1}{\sqrt{2}}
(A\gamma_{5}\Gamma_{5}
+iB\Gamma_{5})
\end{equation}
where we used $\phi=(A+iB)/\sqrt{2}$: In the coupling of $A$ we 
have $(C\gamma_{5}\Gamma_{5})^{T}=-C\gamma_{5}\Gamma_{5}$ but the
difference operator appearing in $\Gamma_{5}$ does not commute 
with $A(x)$, and in the coupling of the 
pseudscalar $B(x)$ we have $(C\Gamma_{5})^{T}\neq -C\Gamma_{5}$.

To cope with this conflict between  lattice chiral symmetry and
the Majorana condition of Yukawa couplings, we choose to impose
the lattice chiral symmetry and analyze the theory
defined by the path integral
\begin{eqnarray}
&&Z=\frac{1}{Z_{0}}\int{\cal D}\phi{\cal D}\phi^{\dagger}
{\cal D}F{\cal D}F^{\dagger}\sqrt{\det[D+m\gamma_{5}\Gamma_{5}
+2g(P_{+}\phi\hat{P}_{+}+P_{-}\phi^{\dagger}\hat{P}_{-})]}
\nonumber\\
&&\times\exp{\{\int-\phi^{\dagger}D^{\dagger}D\phi
+F^{\dagger}F+m[F\Gamma\phi+(F\Gamma\phi)^{\dagger}]
+g[F\Gamma\phi^{2}+(F\Gamma\phi^{2})^{\dagger}]\}}\nonumber\\
\end{eqnarray}
with the normalization factor 
\begin{eqnarray}
Z_{0}&\equiv&\int{\cal D}\phi{\cal D}\phi^{\dagger}
{\cal D}F{\cal D}F^{\dagger}\sqrt{\det[D+m\gamma_{5}\Gamma_{5}}]
\nonumber\\
&&\times\exp\{{\int-\phi^{\dagger}D^{\dagger}D\phi
+F^{\dagger}F+m[F\Gamma\phi+(F\Gamma\phi)^{\dagger}]}\}
\end{eqnarray}
which cancels the divergence coming from the positive 
definite term $F^{\dagger}F$. The square root in (2.17) does 
not define the conventional Pfaffian\footnote{One may note that 
there does not  exist a perfect Euclidean Majorana fermion to 
begin with. } for a finite lattice spacing $a$ because of 
Yukawa couplings. We defined the bosonic part of the action 
in terms of fermionic operators.
The operator $D^{\dagger}D=H^{2}/a^{2}$ is proportional to
a $4\times 4$ unit matrix, but we adopt a convention to 
discard the unit matrix when $D^{\dagger}D=DD^{\dagger}$
appears in the bosonic part of the Lagrangian. In the action 
(2.17) we multiplied the auxiliary field 
$F$ in the mass and interaction terms by an extra factor $\Gamma$
with
\begin{equation}
\Gamma\equiv\sqrt{1-H^{4k+2}}
\end{equation}
to maintain the degeneracy between fermion and boson masses.
This expression of $\Gamma$ is formally equal to 
$\sqrt{\Gamma^{2}_{5}}$,
but we again adopt the convention that $\Gamma^{2}$,unlike 
$\Gamma^{2}_{5}$, does not carry a $4\times 4$ unit Dirac 
matrix. 

In the perturbative treatment, one can reproduce a result 
 equivalent to (2.17) by analyzing a theory defined by a Dirac 
fermion
\begin{eqnarray}
{\cal L}&=&\bar{\psi}D\psi + m\bar{\psi}\gamma_{5}\Gamma_{5}\psi
+ 2g\bar{\psi}(P_{+}\phi\hat{P}_{+}
+P_{-}\phi^{\dagger}\hat{P}_{-})\psi\\
&-&\phi^{\dagger}D^{\dagger}D\phi+F^{\dagger}F
+m[F\Gamma\phi+(F\Gamma\phi)^{\dagger}]
+g[F\Gamma\phi^{2}+(F\Gamma\phi^{2})^{\dagger}]\nonumber  
\end{eqnarray} 
provided that  we adopt the following
calculational rules: We divide by a factor of 2
whenever we encounter a fermion loop contribution in 
perturbation theory. The  quantum effects on the fermion 
lines are treated in the ordinary way. 
The propagators are given by
\begin{eqnarray}
&&\langle\phi\phi^{\dagger}\rangle=\frac{a^{2}}
{H^{2}+(am\Gamma)^{2}}\nonumber\\
&&\langle F F^{\dagger}\rangle=(-)\frac{H^{2}}
{H^{2}+(am\Gamma)^{2}}\nonumber\\
&&\langle F \phi\rangle
=\langle F^{\dagger} \phi^{\dagger}\rangle=(-)\frac{a^{2}m\Gamma}
{H^{2}+(am\Gamma)^{2}}\nonumber\\
&&\langle\psi\bar{\psi}\rangle
=\frac{-1}{D+m\gamma_{5}\Gamma_{5}}=(-)
\frac{a(H+am\Gamma_{5})}{H^{2}+(am\Gamma_{5})^{2}}\gamma_{5}
\end{eqnarray}
and other propagators vanish. When we have $H^{2}$ in 
the bosonic propagators, we adopt the convention to discard
 the unit Dirac matrix in $H^{2}$.

A justification for our choice of the Lagrangian (2.17), in 
particular for our choice of mass terms,  is that 
the free Lagrangian after eliminating the auxiliary field
(written in the original Majorana fermion )
\begin{eqnarray}
{\cal L}_{0}&=&\frac{1}{2}\chi^{T}CD\chi 
+ \frac{1}{2}m\chi^{T}C\gamma_{5}\Gamma_{5}\chi\\
&-&\frac{1}{2}[AD^{\dagger}DA+BD^{\dagger}DB]
-\frac{1}{2}m^{2}[A\Gamma^{2}A+B\Gamma^{2}B] \nonumber\\
&=&\frac{1}{2}\chi^{T}C\gamma_{5}[\frac{1}{a}H+m\Gamma_{5}]\chi\\
&-&\frac{1}{2a^{2}}[AH^{2}A+BH^{2}B]
-\frac{1}{2}m^{2}[A\Gamma^{2}A+B\Gamma^{2}B]
\end{eqnarray}
is invariant under a ``SUSY-like transformation''
\begin{eqnarray}
&&\delta\chi=(\frac{1}{a}H+m\Gamma_{5})\gamma_{5}(A-i\gamma_{5}B)
\epsilon,
\nonumber\\
&&\delta A=\epsilon^{T}C\chi,\nonumber\\
&&\delta B=-i\epsilon^{T}C\gamma_{5}\chi
\end{eqnarray}
with a Majorana parameter $\epsilon$. 
The cancellation of fermionic and bosonic determinants 
for the free Lagrangian is also confirmed by noting
\begin{equation}
\sqrt{\det[H/a+m\Gamma_{5}]}
=\{\det[H/a+m\Gamma_{5}]^{2}\}^{1/4}
=\{\det[H^{2}/a^{2}+m^{2}\Gamma^{2}_{5}]\}^{1/4}
\end{equation}
and the fact that $H^{2}$ and $\Gamma^{2}_{5}$ are proportional
to a $4\times 4$ unit matrix. 
But the symmetry of the full action is not maintained.
See Refs.\cite{kikukawa}\cite{bietenholz}
for discussions of the  related issue.

In passing, we note that the fermion propagator 
in a Majorana notation is given by
\begin{equation}
\langle\chi(y)\chi^{T}(x)C\rangle=(-)\frac{a}{H+am\Gamma_{5}}
\gamma_{5}.
\end{equation}

\subsubsection{Majorana fermion with continuum chiral symmetry
for Yukawa couplings (1)}

We next examine the Lagrangian (2.5) with naive chiral symmetry 
for mass and Yukawa terms but no lattice chiral symmetry . 
After the Majorana reduction, one has
\begin{eqnarray}
{{\cal L}_{M}}_{(2)}&=&\frac{1}{2}\chi^{T}CD\chi 
+ \frac{1}{2}m\chi^{T}C\chi 
+ g\chi^{T}C(P_{+}\phi P_{+}
+P_{-}\phi^{\dagger}P_{-})\chi\nonumber\\
&-&\phi^{\dagger}D^{\dagger}_{1}D_{1}\phi
+F^{\dagger}F
+F(m+D_{2})\phi+(F(m+D_{2})\phi)^{\dagger}\nonumber\\
&&+g[F\phi^{2}+(F\phi^{2})^{\dagger}]\nonumber\\
&=&\frac{1}{2}\chi^{T}CD_{1}\chi 
+ \frac{1}{2}m\chi^{T}C\chi 
+ g\chi^{T}C(P_{+}\phi P_{+}
+P_{-}\phi^{\dagger}P_{-})\chi\nonumber\\
&-&\phi^{\dagger}D^{\dagger}_{1}D_{1}\phi
+F^{\dagger}F
+mF\phi+(mF\phi)^{\dagger}
+g[F\phi^{2}+(F\phi^{2})^{\dagger}]\nonumber\\
&&+\frac{1}{2}\chi^{T}CD_{2}\chi+FD_{2}\phi
+(FD_{2}\phi)^{\dagger}.
\end{eqnarray}
The Majorana reduction works naturally, but we separated the 
kinetic term of the fermion into two parts
\begin{eqnarray}
&&D_{1}\equiv\frac{1}{2}(D-D^{\dagger})=\frac{1}{2a}
(\gamma_{5}H-H\gamma_{5})\propto \pslash,
\nonumber\\
&&D_{2}\equiv\frac{1}{2}(D+D^{\dagger})=\frac{1}{2a}
(\gamma_{5}H+H\gamma_{5})=\frac{1}{a}H^{2k+2}
\end{eqnarray}
to ensure the degeneracy of fermion and boson mass spectrum
and also the one-loop tadpole cancellation.
Here we used the fact that $D_{2}$ is independent of Dirac 
matrices for the free fermion case.
Note that in general
\begin{eqnarray}
&&\{\gamma_{5},D_{1}\}=\frac{1}{2a}\{\gamma_{5},
(\gamma_{5}H-H\gamma_{5}) \}=0,\nonumber\\
&&[\gamma_{5}, D_{2}]=\frac{1}{2a}[\gamma_{5}, 
(\gamma_{5}H+H\gamma_{5})]=0.
\end{eqnarray}
In the Lagrangian (2.28), the kinetic (K\"{a}hler) term is given 
by
\begin{equation}
{{{\cal L}_{M}}_{(2)}}_{kin}=
\frac{1}{2}\chi^{T}CD_{1}\chi 
-\phi^{\dagger}D^{\dagger}_{1}D_{1}\phi
+F^{\dagger}F
\end{equation}
and a part of the Ginsparg-Wilson operator is transferred 
to a potential part (which however contains derivative operators)
\begin{equation}
{{{\cal L}_{M}}_{(2)}}_{pot}=
\frac{1}{2}\chi^{T}CD_{2}\chi+FD_{2}\phi+(FD_{2}\phi)^{\dagger}
\end{equation}
which is treated in a manner analogous to the Wilson term in the 
conventional lattice formulation. This potential term induces a 
hard breaking of continuum chiral symmetry.

The propagators are given by
\begin{eqnarray}
\langle\phi\phi^{\dagger}\rangle&=&\frac{1}
{D_{1}^{\dagger}D_{1}+(m+D_{2})^{2}}\nonumber\\
&=&\frac{a^{2}}
{H^{2}+ma(\gamma_{5}H+H\gamma_{5})+(am)^{2}}\nonumber\\
&=&\frac{a^{2}}
{H^{2}+2maH^{2k+2}+(am)^{2}},\nonumber\\
\langle F F^{\dagger}\rangle&=&(-)\frac{a^{2}D_{1}^{\dagger}D_{1}
 }
{H^{2}+2maH^{2k+2}+(am)^{2}}\nonumber\\
&=&(-)\frac{H^{2}-H^{4k+4}}
{H^{2}+2maH^{2k+2}+(am)^{2}},\nonumber\\
\langle F \phi\rangle
&=&\langle F^{\dagger} \phi^{\dagger}\rangle
=(-)\frac{a^{2}(m+D_{2})}
{H^{2}+2maH^{2k+2}+(am)^{2}}\nonumber\\
&=&(-)\frac{a(am+H^{2k+2})}
{H^{2}+2maH^{2k+2}+(am)^{2}},\nonumber\\
\langle\chi\chi^{T}C\rangle
&=&\frac{-1}{D+m}=(-)\frac{a}{H+am\gamma_{5}}\gamma_{5}\nonumber\\
&=&(-)
\frac{a(H+am\gamma_{5})}
{H^{2}+2maH^{2k+2}+(am)^{2}}\gamma_{5}
\end{eqnarray}
and other propagators vanish.

This scheme corresponds to the analysis of Bartels and Kramer
\cite{bartels} on the basis of the conventional Wilson fermion, 
if one identifies $D_{1}$ with the naive lattice Dirac
operator and $D_{2}$ as the Wilson term. This decomposition
crucially depends on the fact that $D_{2}$ is independent of 
Dirac matrices, which is not the case in the presence of 
background gauge field, for example. (In such a case, one need to
define $D_{2}$ in the bosonic sector independently from $D_{2}$
in the fermionic sector.) In this sense, 
the Ginsparg-Wilson operator is less flexible than 
the naive lattice fermion with a Wilson term\cite{bartels}
 when applied to a scalar multiplet.

\subsubsection{Majorana fermion with continuum chiral symmetry
for Yukawa couplings (2)}

We next examine the Lagrangian (2.7) with naive chiral symmetry 
for mass and Yukawa terms but with an extra factor of $\gamma_{5}$
. After the Majorana reduction, one has
\begin{eqnarray}
{{\cal L}_{M}}_{(3)}&=&\frac{1}{2}\chi^{T}CD\chi 
+ \frac{i}{2}m\chi^{T}C\gamma_{5}\chi 
+ ig\chi^{T}C\gamma_{5}(P_{+}\phi P_{+}
+P_{-}\phi^{\dagger}P_{-})\chi\nonumber\\
&-&\phi^{\dagger}D^{\dagger}D\phi
+F^{\dagger}F
+mF\phi+m(F\phi)^{\dagger}
+g[F\phi^{2}+(F\phi^{2})^{\dagger}].
\end{eqnarray}
The Majorana reduction works naturally, and 
the propagators are given by
\begin{eqnarray}
\langle\phi\phi^{\dagger}\rangle&=&\frac{1}
{D^{\dagger}D+m^{2}}\nonumber\\
&=&\frac{a^{2}}
{H^{2}+(am)^{2}}\nonumber\\
\langle F F^{\dagger}\rangle&=&(-)\frac{H^{2}}
{H^{2}+(am)^{2}}\nonumber\\
\langle F \phi\rangle
&=&\langle F^{\dagger} \phi^{\dagger}\rangle
=(-)\frac{a^{2}m}
{H^{2}+(am)^{2}}\nonumber\\
\langle\chi\chi^{T}C\rangle
&=&\frac{-1}{D+im\gamma_{5}}
=(-)\frac{a}{H+iam}\gamma_{5}\nonumber\\
&=&(-)
\frac{a(H-iam)}
{H^{2}+(am)^{2}}\gamma_{5}
\end{eqnarray}
and other propagators vanish.  

This scheme is the simplest in many respects, but this scheme 
leads to non-vanishing  fermion tadpole contributions for both 
of the scalar $A$ and the pseudscalar $B$.  This means that the 
vacuum breaks CP symmetry. To be precise, the one-loop fermion
tadpole diagram is evaluated by
\begin{eqnarray}
&&\frac{1}{2}Tr\frac{1}{D+im\gamma_{5}}
2ig\gamma_{5}(P_{+}\phi P_{+}
+P_{-}\phi^{\dagger}P_{-})\nonumber\\
&&=\frac{2iga}{2}Tr(P_{+}\phi
+P_{-}\phi^{\dagger})\frac{H-ima}{H^{2}
+(ma)^{2}}\nonumber\\
&&=\frac{2iga}{4}
Tr[(\phi+\phi^{\dagger})+(\phi-\phi^{\dagger})\gamma_{5}]
\frac{H-ima}{H^{2}
+(ma)^{2}}\nonumber\\
&&=\frac{2ga}{4}(\phi+\phi^{\dagger})
Tr\frac{iH+ma}{H^{2}+(ma)^{2}}\nonumber\\
&&+\frac{2iga}{4}
(\phi-\phi^{\dagger})Tr\frac{\gamma_{5}H}{H^{2}+(ma)^{2}}
\nonumber\\
&&=\frac{\sqrt{2}gma^{2}}{2}A
Tr\frac{1}{H^{2}+(ma)^{2}}\nonumber\\
&&-\frac{\sqrt{2}ga}{2}B
Tr\frac{H^{2k+2}}{H^{2}+(ma)^{2}}\nonumber\\
&&=\frac{\sqrt{2}gm}{2a^{2}}A
tr\int^{\pi}_{-\pi}\frac{d^{4}k}{(2\pi)^{4}}
\frac{1}{H^{2}(k)+(ma)^{2}}
\nonumber\\
&&-\frac{\sqrt{2}g}{2a^{3}}B
tr\int^{\pi}_{-\pi}\frac{d^{4}k}{(2\pi)^{4}}
\frac{H^{2k+2}(k)}{H^{2}(k)+(ma)^{2}}
\end{eqnarray}
where we used the fact that $H^{2}$ is independent of Dirac 
matrices and $\gamma_{5}H+H\gamma_{5}=2H^{2k+2}$.
We also rescaled the loop momentum 
$ak_{\mu}\rightarrow k_{\mu}$,
and the remaining trace is over Dirac indices.
 The tadpole for 
the scalar $A$ is quadratically divergent and it is cancelled 
by a scalar one-loop tadpole contribution. On the other hand,
the tadpole for the pseudscalar $B$ gives cubic 
divergence, and it 
cannot be cancelled by scalar loop diagrams unless one introduces 
complicated CP violating interactions among $\phi$ and $F$.

In passing we note that our scheme, which uses the scalar 
density (1.18) as a mass term, provides a viable alternative to 
a Majorana representation of a (massive) chiral gauge theory with 
a real representation of gauge group discussed in\cite{suzuki}, 
namely
\begin{equation}
{\cal L}=\frac{1}{2}\chi^{T}CD\chi 
+ \frac{1}{2}m\chi^{T}C\gamma_{5}\Gamma_{5}\chi
\end{equation}
if one includes gauge field in $D$ and adds a Lagrangian
for the gauge field. The fermion propagator in this case is 
given by (2.27).

\section{Weyl Representation and Holomorphic Properties}

We have so far examined the lattice Majorana representation.
We now discuss the lattice Weyl representation, which is 
more fundamental from a view point of supersymmetry\footnote{
In continuum (and Minkowski) theory the equivalence of the 
Majorana and Weyl representations is trivially valid, but on the 
lattice their equivalence is not evident.}.
An analysis of weyl fermion with a real representation of 
gauge group suggests the following Lagrangian\cite{suzuki}
\begin{equation}
{\cal L}^{(0)}_{(4)}=\bar{\psi}_{L}D\psi_{L}
+\frac{1}{2}im\psi_{L}^{T}B\psi_{L}
-\frac{1}{2}im\bar{\psi}_{L}B^{-1}\bar{\psi}^{T}_{L}.
\end{equation}
Here $B=C\gamma_{5}$, and this expression is consistent if one
recalls $\psi_{L}=\hat{P}_{-}\psi$, 
$\bar{\psi}_{L}=\bar{\psi}P_{+}$ and the properties
\begin{eqnarray}
B\hat{P}_{-}&=&C\gamma_{5}\frac{1}{2}(1-\hat{\gamma}_{5})
=C\gamma_{5}\frac{1}{2}(1+\gamma_{5}-2\Gamma_{5})\nonumber\\
&=&-(C\gamma_{5}\frac{1}{2}(1+\gamma_{5}-2\Gamma_{5}))^{T}
=-\hat{P}^{T}_{-}B^{T}=\hat{P}^{T}_{-}B,\nonumber\\
P_{+}B^{-1}&=&B^{-1}BP_{+}B^{-1}=B^{-1}P^{T}_{+}
\end{eqnarray}
where we used $B^{T}=-B$.
Note that if one replaces $B$ by $C$
in (3.1), it does not work since the first relation in (3.2) 
does not hold; $C\Gamma_{5}\neq -(C\Gamma_{5})^{T}$. 
We adopt the Lagrangian (3.1) as a basis of our analysis since 
we have a free Dirac operator $D$.

A natural way to introduce the Yukawa coupling is to consider
\begin{eqnarray}
{{\cal L}_{L}}_{(4)}&=&\bar{\psi}_{L}D\psi_{L}
+\frac{1}{2}im\psi_{L}^{T}B\psi_{L}
-\frac{1}{2}im\bar{\psi}_{L}B^{-1}\bar{\psi}^{T}_{L}\nonumber\\
&&+ig\psi_{L}^{T}B\phi\psi_{L}
-ig\bar{\psi}_{L}B^{-1}\phi^{\dagger}\bar{\psi}^{T}_{L}
\\
&&-\phi^{\dagger}D^{\dagger}D\phi
+F^{\dagger}F
+mF\phi+m(F\phi)^{\dagger}
+g[F\phi^{2}+(F\phi^{2})^{\dagger}]\nonumber
\end{eqnarray}
which is satisfactory from a view point of holomorphic 
properties in the sense that the potential terms consist of
either $(\phi,\psi_{L},F )$ or 
$(\phi^{\dagger},\bar{\psi}_{L},F^{\dagger})$, and these sets 
of fields mix only in the kinetic terms
\begin{eqnarray}
{{{\cal L}_{L}}_{(4)}}_{kin}&=&\bar{\psi}_{L}D\psi_{L}
-\phi^{\dagger}D^{\dagger}D\phi+F^{\dagger}F\nonumber\\
&=&\bar{\psi}_{L}\frac{H}{a}\psi_{L}
-\phi^{\dagger}(\frac{H}{a})^{2}\phi+F^{\dagger}F.
\end{eqnarray}
The mass terms are also generated by a shift 
$\phi\rightarrow \phi+m/(2g)$ in the interaction terms. 

However the Yukawa coupling in this formula (3.3) is not 
consistent as a Weyl decomposition. This fact is seen as 
follows: One may first define
a Lagrangian in terms of unconstrained variables $\psi$ and 
$\bar{\psi}$ as 
\begin{eqnarray}
{\cal L}_{(4)}&=&\bar{\psi}D\psi
+\frac{1}{2}im\psi^{T}B\psi
-\frac{1}{2}im\bar{\psi}B^{-1}\bar{\psi}^{T}\nonumber\\
&&+ig\psi^{T}B\phi\psi
-ig\bar{\psi}\phi^{\dagger}B^{-1}\bar{\psi}^{T}
\\
&&-\phi^{\dagger}D^{\dagger}D\phi
+F^{\dagger}F
+mF\phi+m(F\phi)^{\dagger}
+g[F\phi^{2}+(F\phi^{2})^{\dagger}]\nonumber
\end{eqnarray}
and define the path integral
\begin{equation}
\int{\cal D}\psi{\cal D}\bar{\psi}\exp[\int {\cal L}_{(4)}].
\end{equation}
A Weyl decomposition means to rewrite this path integral as 
a product of left and right components as  
( for the moment, by paying attention only to the fermion 
sector)
\begin{equation}
\int{\cal D}\psi{\cal D}\bar{\psi}\exp[\int {\cal L}_{(4)}]
=\int{\cal D}\psi_{L}{\cal D}\bar{\psi}_{L}
\exp[\int {{\cal L}_{L}}_{(4)}]
\int{\cal D}\psi_{R}{\cal D}\bar{\psi}_{R}
\exp[\int {{\cal L}_{R}}_{(4)}].
\end{equation}
But this fails since we have a cross term in the Yukawa 
coupling
\begin{eqnarray}
&&ig \int[\psi_{L}^{T}B\phi\psi_{R}+\psi_{R}^{T}B\phi\psi_{L}] 
\nonumber\\
&&=2ig \int\psi_{R}^{T}B\phi\psi_{L}
=2ig \int\psi_{R}^{T}B\phi\hat{P}_{-}\psi\nonumber\\
&&=2ig \int\psi_{R}^{T}\phi\hat{P}^{T}_{-}B\psi
=2ig \int\psi_{R}^{T}[\phi,\hat{P}^{T}_{-}]B\psi\nonumber\\
&&=2ig \int\psi_{R}^{T}[\phi,(H^{2k+1})^{T}]B\psi\neq 0
\end{eqnarray}
where we used $B^{T}=-B$, $\psi_{R}^{T}\hat{P}^{T}_{-}=0$ and 
an explicit expression of $\hat{P}^{T}_{-}$.
For the unconstrained variable $\psi$, the vanishing 
condition of this cross term requires
\begin{equation}
\psi_{R}^{T}[\phi(x),(H^{2k+1})^{T}]=0
\end{equation}
which however does not hold\footnote{Alternative way to see this 
complication is to note that the interaction term 
$ig\int\psi_{L}^{T}B\phi\psi_{L}$ in (3.3) does not satisfy the 
consistency condition in the sense of (3.2) because of the 
presence of $\phi(x)$. The field $\phi(x)$ connects $\psi_{L}$
to $\psi_{R}^{T}$ as in (3.8).}, since the 
difference operator 
appearing in $H$ does not commute with $\phi(x)$.

In view of the absence of the precise Weyl decomposition
in the present framework, one may use the Lagrangian (3.5)
and divide all the fermion loop diagrams by a factor of $2$ in 
perturbative calculations.
 The propagators in (3.5) are given by
\begin{eqnarray}
\langle \psi\bar{\psi}\rangle&=&\frac{-1}{D^{\dagger}D+m^{2}}
D^{\dagger}=\frac{-a}{H^{2}+(am)^{2}}H\gamma_{5}
\nonumber\\
\langle \psi\psi^{T}B\rangle&=&\frac{im}{D^{\dagger}D+m^{2}}
=\frac{ima^{2}}{H^{2}+(am)^{2}}\nonumber\\
\langle B^{-1}\bar{\psi}^{T}\bar{\psi}\rangle&
=&\frac{-im}{D^{\dagger}D+m^{2}}=\frac{-ima^{2}}{H^{2}+(am)^{2}}
\nonumber\\
\langle\phi\phi^{\dagger}\rangle&=&\frac{1}
{D^{\dagger}D+m^{2}}=\frac{a^{2}}{H^{2}+(am)^{2}}\nonumber\\
\langle F F^{\dagger}\rangle&=&(-)\frac{H^{2}}
{H^{2}+(am)^{2}}\nonumber\\
\langle F \phi\rangle
&=&\langle F^{\dagger} \phi^{\dagger}\rangle
=(-)\frac{a^{2}m}
{H^{2}+(am)^{2}}.
\end{eqnarray}
For the Lagrangian (3.5), one can confirm that no CP violation 
appears in one-loop level fermion tadpoles and that the fermion 
tadpoles are precisely cancelled by scalar tadpoles. The 
Lagrangian (3.5) preserves a certain aspect of holomorphic 
properties but 
the information of lattice chiral symmetry is lost in 
Yukawa couplings\footnote{This property is somewhat 
complementary to (2.20), where lattice chiral symmetry is 
manifest but holomorphic properties are not obvious.}; 
consequently, the potential part
including fermion mass terms retains a good renormalization
property but the fermion self-energy part induces a linear 
divergence (and thus a Dirac mass ). Incidentally, if the chiral 
decomposition (3.7) should be exact, no linear divergence would 
appear in the one-loop fermion self-energy for (3.5).

We thus recognize a conflict between the lattice chiral
symmetry and Yukawa couplings for the Weyl representation also.
An alternative Weyl representation is obtained by
rewriting ${{\cal L}_{M}}_{(2)}$ in (2.28) in terms of 
two-component 
spinors. In this sense, only the Lagrangian (2.28) is 
consistent with the Majorana or Weyl condition, but of course 
(2.28) explicitly breaks lattice chiral symmetry by Yukawa
couplings from the beginning; besides the term 
${{{\cal L}_{M}}_{(2)}}_{pot}$
in (2.32) breaks $U(1)$ in $U(1)\times U(1)_{R}$ symmetry in 
the kinetic (K\"{a}hler) term ${{{\cal L}_{M}}_{(2)}}_{kin}$ 
(2.31) 
even for $m=g=0$; here $U(1)_{R}$ stands for R-symmetry.

\section{One-loop Perturbative Analysis}

In this Section we examine in some detail the lattice 
regularization of the 
Wess-Zumino model (2.17) which preserves lattice chiral symmetry. 
We  show that the corrections to all the interaction 
terms (including mass terms ) of all the field 
variables $\psi$ and $\phi$ vanish in the one-loop level for a 
{\em finite} lattice spacing $a$ when 
renormalized  at vanishing momenta. This 
in particular shows that the quadratic divergences in the mass
term of the scalar particle are exactly cancelled in the 
one-loop level.
In this sense, our regularization preserves a certain essential
aspect of the supersymmetric model.

\subsection{Fermionic One-loop Contributions}

The one-loop fermion contribution is evaluated by
\begin{eqnarray}
&&\sqrt{\det[D+m\gamma_{5}\Gamma_{5}+2g(P_{+}\phi\hat{P}_{+}
+P_{-}\phi^{\dagger}\hat{P}_{-})]}
\nonumber\\
&&=\exp\{\frac{1}{2}Tr\ln[D+m\gamma_{5}\Gamma_{5}+
2g(P_{+}\phi\hat{P}_{+}
+P_{-}\phi^{\dagger}\hat{P}_{-})]\}
\nonumber\\
&&=\exp\{\frac{1}{2}Tr\ln[D+m\gamma_{5}\Gamma_{5}]
+\frac{1}{2}Tr\frac{1}{D+m\gamma_{5}\Gamma_{5}}
2g(P_{+}\phi\hat{P}_{+}
+P_{-}\phi^{\dagger}\hat{P}_{-})\nonumber\\
&&-\frac{1}{4}
Tr\frac{1}{D+m\gamma_{5}\Gamma_{5}}
2g(P_{+}\phi\hat{P}_{+}
+P_{-}\phi^{\dagger}\hat{P}_{-})\frac{1}{D+m\gamma_{5}\Gamma_{5}}
2g(P_{+}\phi\hat{P}_{+}
+P_{-}\phi^{\dagger}\hat{P}_{-})\nonumber\\
&&+...\}.
\end{eqnarray}

\subsubsection{Tadpole Diagrams}
We first examine the tadpole terms, namely, terms linear in 
$\phi$ and $\phi^{\dagger}$.
 We have
\begin{eqnarray}
&&\frac{1}{2}Tr\frac{1}{D+m\gamma_{5}\Gamma_{5}}
2g(P_{+}\phi\hat{P}_{+}
+P_{-}\phi^{\dagger}\hat{P}_{-})\nonumber\\
&&=\frac{2ga}{2}Tr(P_{+}\phi\Gamma_{5}
+P_{-}\phi^{\dagger}\Gamma_{5})\frac{H+ma\Gamma_{5}}{H^{2}
+(ma\Gamma_{5})^{2}}\nonumber\\
&&=\frac{2gma^{2}}{2}
Tr(P_{+}\phi
+P_{-}\phi^{\dagger})\frac{\Gamma^{2}_{5}}{H^{2}
+(ma\Gamma_{5})^{2}}\nonumber\\
&&=\frac{2gma^{2}}{4}
(\phi+\phi^{\dagger})Tr\frac{\Gamma^{2}_{5}}{H^{2}
+(ma\Gamma_{5})^{2}}
\end{eqnarray}
where we used $P_{\pm}\Gamma_{5}HP_{\pm}=0$ and the fact that 
$H^{2}$ and $\Gamma^{2}_{5}$
are independent of $\gamma^{\mu}$ matrices. The trace for  
fermion loop amplitudes includes the integral over internal 
momentum as well as the sum over Dirac indices, and the above
integral is quadratically divergent\footnote{The power of
divergence in the limit $a\rightarrow 0$ is read from the powers 
of $a$ appearing in front of these expressions. For example,
$a^{2}$ corresponds to 
quadratic divergence and $a^{4}$ corresponds to logarithmic 
divergence. }. We also set the
momenta carried by external fields at $0$. 
Note that only the scalar component $\phi+\phi^{\dagger}=
\sqrt{2}A$ develops the vacuum value.
The trace over 
Dirac matrices gives an extra factor of $4$ when compared to
the scalar contribution.  

\subsubsection{Self-energy Correction for Scalar Particles}

The self-energy corrections of scalar particles by fermion loop 
diagrams are given by
\begin{eqnarray}
&&-\frac{1}{4}
Tr\frac{1}{D+m\gamma_{5}\Gamma_{5}}
2g(P_{+}\phi\hat{P}_{+}
+P_{-}\phi^{\dagger}\hat{P}_{-})\frac{1}{D+m\gamma_{5}\Gamma_{5}}
2g(P_{+}\phi\hat{P}_{+}
+P_{-}\phi^{\dagger}\hat{P}_{-})\nonumber\\
&&=-\frac{(2ga)^{2}}{4}
Tr\frac{H+ma\Gamma_{5}}{H^{2}+(ma\Gamma_{5})^{2}}
(P_{+}\phi\Gamma_{5}
+P_{-}\phi^{\dagger}\Gamma_{5})
\frac{H+ma\Gamma_{5}}{H^{2}+(ma\Gamma_{5})^{2}}
(P_{+}\phi\Gamma_{5}
+P_{-}\phi^{\dagger}\Gamma_{5})\nonumber\\
&&=-\frac{(2ga)^{2}}{4}\nonumber\\
&&\times[2TrP_{-}\phi^{\dagger}
\frac{\Gamma_{5}H}{H^{2}+(ma\Gamma_{5})^{2}}
P_{+}\phi
\frac{\Gamma_{5}H}{H^{2}+(ma\Gamma_{5})^{2}}\nonumber\\
&&+TrP_{+}\phi
\frac{ma\Gamma^{2}_{5}}{H^{2}+(ma\Gamma_{5})^{2}}
P_{+}\phi
\frac{ma\Gamma^{2}_{5}}{H^{2}+(ma\Gamma_{5})^{2}}\nonumber\\
&&+TrP_{-}\phi^{\dagger}
\frac{ma\Gamma^{2}_{5}}{H^{2}+(ma\Gamma_{5})^{2}}
P_{-}\phi^{\dagger}
\frac{ma\Gamma^{2}_{5}}{H^{2}+(ma\Gamma_{5})^{2}}]\nonumber\\
&&=-\frac{(2ga)^{2}}{4}\nonumber\\
&&\times[-Tr\phi^{\dagger}
\frac{H\Gamma_{5}}{H^{2}+(ma\Gamma_{5})^{2}}
\phi\frac{\Gamma_{5}H}{H^{2}+(ma\Gamma_{5})^{2}}\nonumber\\
&&+\frac{(ma)^{2}}{2}Tr\phi
\frac{\Gamma^{2}_{5}}{H^{2}+(ma\Gamma_{5})^{2}}
\phi\frac{\Gamma^{2}_{5}}{H^{2}+(ma\Gamma_{5})^{2}}\nonumber\\
&&+\frac{(ma)^{2}}{2}Tr\phi^{\dagger}
\frac{\Gamma^{2}_{5}}{H^{2}+(ma\Gamma_{5})^{2}}
\phi^{\dagger}\frac{\Gamma^{2}_{5}}{H^{2}+(ma\Gamma_{5})^{2}}]
\end{eqnarray}
where we used $P_{\pm}\Gamma_{5}H=-H\Gamma_{5}P_{\mp}$,
 $P_{\pm}\Gamma_{5}HP_{\pm}=0$ and 
$P_{\pm}\Gamma^{2}_{5}P_{\mp}=0$.
The first term is quadratically divergent, and the last two 
terms are logarithmically divergent. When one sets the 
external
momentum at $0$ to define the mass renormalization factor, these
expressions are written as
\begin{eqnarray}
&&\frac{(2ga)^{2}}{4}\phi^{\dagger}\phi
Tr[\frac{\Gamma^{2}_{5}}{H^{2}+(ma\Gamma_{5})^{2}}]\\
&&-\frac{(2ga)^{2}(ma)^{2}}{8}(2\phi^{\dagger}\phi
+\phi^{\dagger}\phi^{\dagger}+\phi\phi) Tr[
\frac{\Gamma^{2}_{5}}{H^{2}+(ma\Gamma_{5})^{2}}
\frac{\Gamma^{2}_{5}}{H^{2}+(ma\Gamma_{5})^{2}}]\nonumber
\end{eqnarray}
where the first term gives subtractive  renormalization 
and the remaining 3 terms give multiplicative renormalization
in the conventional classification. But these terms do not appear
in the bare Lagrangian with the auxiliarly field $F$ (2.17) and 
(2.20), and 
they should be precisely cancelled by scalar contributions. 
To discuss the cancellation of these terms by scalar
loop contributions, we need to take account of an extra factor 
of $4$ arising from the trace over Dirac matrices. 

\subsubsection{Triple Couplings of Scalars}

The fermion contribution to triple couplings
is given by
\begin{eqnarray}
&&\frac{1}{6}Tr\{ 
\frac{a}{H+ma\Gamma_{5}}\gamma_{5}2g[P_{+}\phi\hat{P}_{+}
+P_{-}\phi^{\dagger}\hat{P}_{-}]\nonumber\\
&&\times \frac{a}{H+ma\Gamma_{5}}\gamma_{5}2g[P_{+}\phi\hat{P}_{+}
+P_{-}\phi^{\dagger}\hat{P}_{-}]\nonumber\\
&&\times \frac{a}{H+ma\Gamma_{5}}\gamma_{5}2g[P_{+}\phi\hat{P}_{+}
+P_{-}\phi^{\dagger}\hat{P}_{-}]\}\nonumber\\
&&=\frac{4(ga)^{3}}{3}Tr\{
\frac{H+ma\Gamma_{5}}{H^{2}+(ma\Gamma_{5})^{2}}
[P_{+}\phi\Gamma_{5}
+P_{-}\phi^{\dagger}\Gamma_{5}]\nonumber\\
&&\times\frac{H+ma\Gamma_{5}}{H^{2}+(ma\Gamma_{5})^{2}}
[P_{+}\phi\Gamma_{5}
+P_{-}\phi^{\dagger}\Gamma_{5}]\nonumber\\
&&\times\frac{H+ma\Gamma_{5}}{H^{2}+(ma\Gamma_{5})^{2}}
[P_{+}\phi\Gamma_{5}
+P_{-}\phi^{\dagger}\Gamma_{5}]\}.
\end{eqnarray}
By using the relations
$P_{\pm}\Gamma_{5}H=-H\Gamma_{5}P_{\mp}$ and 
$P_{\pm}\Gamma_{5}HP_{\pm}=0$,
we have, for example,
\begin{eqnarray}
&&\frac{4(ga)^{3}}{3}Tr\{\phi P_{+}\Gamma_{5}
\frac{H+ma\Gamma_{5}}{H^{2}+(ma\Gamma_{5})^{2}}
\phi P_{+}\Gamma_{5}\frac{H+ma\Gamma_{5}}
{H^{2}+(ma\Gamma_{5})^{2}}\nonumber\\
&&\times
\phi P_{+}\Gamma_{5}\frac{H+ma\Gamma_{5}}
{H^{2}+(ma\Gamma_{5})^{2}}\}\\
&&=\frac{4(ga)^{3}(ma)^{3}}{3}\nonumber\\
&&\times Tr\{\phi P_{+}
\frac{\Gamma^{2}_{5}}{H^{2}+(ma\Gamma_{5})^{2}}
\phi P_{+}\frac{\Gamma^{2}_{5}}
{H^{2}+(ma\Gamma_{5})^{2}}
\phi P_{+}\frac{\Gamma^{2}_{5}}
{H^{2}+(ma\Gamma_{5})^{2}}\}\nonumber
\end{eqnarray}
which is convergent. Similarly, we can confirm that
the coupling $\phi^{\dagger}\phi^{\dagger}\phi^{\dagger}$
is convergent.\\

We next examine
\begin{eqnarray}
&&\frac{4(ga)^{3}}{3}Tr\{\phi^{\dagger} P_{-}\Gamma_{5}
\frac{H+ma\Gamma_{5}}{H^{2}+(ma\Gamma_{5})^{2}}
\phi P_{+}\Gamma_{5}\frac{H+ma\Gamma_{5}}
{H^{2}+(ma\Gamma_{5})^{2}}\nonumber\\
&&\times
\phi P_{+}\Gamma_{5}\frac{H+ma\Gamma_{5}}
{H^{2}+(ma\Gamma_{5})^{2}}\}\\
&&=\frac{4(ga)^{3}ma}{3}\nonumber\\
&&\times Tr\{\phi^{\dagger} P_{-}
\frac{\Gamma_{5}H}{H^{2}+(ma\Gamma_{5})^{2}}
\phi P_{+}\frac{\Gamma^{2}_{5}}
{H^{2}+(ma\Gamma_{5})^{2}}
\phi P_{+}\frac{\Gamma_{5}H}
{H^{2}+(ma\Gamma_{5})^{2}}\}\nonumber
\end{eqnarray}
which is logarithmically divergent. We thus set the external
momenta at $0$ to define the renormalization factor at 
vanishing momenta, and we obtain
\begin{eqnarray}
&&-\frac{4(ga)^{3}ma}{3}\phi^{\dagger}\phi\phi\nonumber\\
&&\times Tr\{
\frac{H\Gamma_{5}}{H^{2}+(ma\Gamma_{5})^{2}}
P_{+}\frac{\Gamma^{2}_{5}}
{H^{2}+(ma\Gamma_{5})^{2}}\frac{\Gamma_{5}H}
{H^{2}+(ma\Gamma_{5})^{2}}\}\nonumber\\
&&=-\frac{4(ga)^{3}ma}{6}\phi^{\dagger}\phi\phi\nonumber\\
&&\times Tr\{
\frac{H^{2}}{H^{2}+(ma\Gamma_{5})^{2}}
\frac{\Gamma^{2}_{5}}
{H^{2}+(ma\Gamma_{5})^{2}}\frac{\Gamma^{2}_{5}}
{H^{2}+(ma\Gamma_{5})^{2}}\}.
\end{eqnarray}
Since we have 3 combinations for this choice, we have the 
numerical coefficient
\begin{equation}
-\frac{4(ga)^{3}ma}{6}\times 3\times 4=-\frac{48(ga)^{3}ma}{6}
\end{equation}
where the last factor of 4 comes from the trace over Dirac
matrices.

We have the same numerical coefficient for the combination
$\phi^{\dagger}\phi^{\dagger}\phi$.

\subsubsection{Quartic Couplings}

The quartic couplings from the fermion contribution are given
by
\begin{eqnarray}
&&-\frac{(2ga)^{4}}{8}Tr\{
\frac{H+ma\Gamma_{5}}{H^{2}+(ma\Gamma_{5})^{2}}
[P_{+}\phi\Gamma_{5}
+P_{-}\phi^{\dagger}\Gamma_{5}]\nonumber\\
&&\times\frac{H+ma\Gamma_{5}}{H^{2}+(ma\Gamma_{5})^{2}}
[P_{+}\phi\Gamma_{5}
+P_{-}\phi^{\dagger}\Gamma_{5}]\nonumber\\
&&\times\frac{H+ma\Gamma_{5}}{H^{2}+(ma\Gamma_{5})^{2}}
[P_{+}\phi\Gamma_{5}
+P_{-}\phi^{\dagger}\Gamma_{5}]\nonumber\\
&&\times\frac{H+ma\Gamma_{5}}{H^{2}+(ma\Gamma_{5})^{2}}
[P_{+}\phi\Gamma_{5}
+P_{-}\phi^{\dagger}\Gamma_{5}]\}.
\end{eqnarray}
It can be confirmed that only the combination of the form
$\phi^{\dagger}\phi\phi^{\dagger}\phi$ is logarithmically
divergent. We thus examine, for example,
\begin{eqnarray}
&&-\frac{(2ga)^{4}}{8}Tr\{P_{-}\phi^{\dagger}\Gamma_{5}
\frac{H+ma\Gamma_{5}}{H^{2}+(ma\Gamma_{5})^{2}}
P_{+}\phi\Gamma_{5}
\frac{H+ma\Gamma_{5}}{H^{2}+(ma\Gamma_{5})^{2}}\nonumber\\
&&\times P_{-}\phi^{\dagger}\Gamma_{5}
\frac{H+ma\Gamma_{5}}{H^{2}+(ma\Gamma_{5})^{2}}
P_{+}\phi\Gamma_{5}
\frac{H+ma\Gamma_{5}}{H^{2}+(ma\Gamma_{5})^{2}}\}\nonumber\\
&&=-\frac{(2ga)^{4}}{8}Tr\{P_{-}\phi^{\dagger}
\frac{\Gamma_{5}H}{H^{2}+(ma\Gamma_{5})^{2}}
P_{+}\phi
\frac{\Gamma_{5}H}{H^{2}+(ma\Gamma_{5})^{2}}\nonumber\\
&&\times P_{-}\phi^{\dagger}
\frac{\Gamma_{5}H}{H^{2}+(ma\Gamma_{5})^{2}}
P_{+}\phi
\frac{\Gamma_{5}H}{H^{2}+(ma\Gamma_{5})^{2}}\}\nonumber\\
&&=-\frac{(2ga)^{4}}{16}\phi^{\dagger}\phi\phi^{\dagger}\phi
\\
&&\times
Tr\{\frac{H^{2}}{H^{2}+(ma\Gamma_{5})^{2}}
\frac{\Gamma^{2}_{5}}{H^{2}+(ma\Gamma_{5})^{2}}
\frac{H^{2}}{H^{2}+(ma\Gamma_{5})^{2}}
\frac{\Gamma^{2}_{5}}{H^{2}+(ma\Gamma_{5})^{2}}\}\nonumber
\end{eqnarray}
where we set the external momenta at $0$ to extract the 
divergent coefficient. We have two contributions of this 
combination, and thus the total coefficient is given by
\begin{equation}
-\frac{(2ga)^{4}}{16}\times 2\times 4=-\frac{(2ga)^{4}}{2}
\end{equation}
where the last factor $4$ is from the trace over Dirac
matrices.

It is confirmed that all the higher order couplings are finite.

\subsection{Scalar Contributions}

We next examine the scalar contributions 
to various scalar couplings on the basis of
the potential 
\begin{equation}
V=g[F\Gamma\phi^{2}+(F\Gamma\phi^{2})^{\dagger}]. 
\end{equation}

\subsubsection{Tadpole Contribution}
The tadpole is evaluated by 
\begin{eqnarray}
&&2g[\langle F\Gamma\phi\rangle\phi+\langle F^{\dagger}
\Gamma\phi^{\dagger}\rangle\phi^{\dagger}]\nonumber\\
&&=-2g\int[\phi\frac{a^{2}m\Gamma^{2}}
{H^{2}+(am\Gamma)^{2}}
+\phi^{\dagger}\frac{a^{2}m\Gamma^{2}}
{H^{2}+(am\Gamma)^{2}}]\nonumber\\
&&=-2mga^{2}(\phi+\phi^{\dagger})\int\frac{\Gamma^{2}}
{H^{2}+(am\Gamma)^{2}}
\end{eqnarray}
which is precisely what we need to cancel the fermion 
contribution (4.2).

\subsubsection{Self-energy for Scalar Particles}

The one-loop self-energy of scalar particles is given by 
\begin{eqnarray}
&&\frac{g^{2}}{2!}[F\Gamma\phi^{2}
+(F\Gamma\phi^{2})^{\dagger}]
[F\Gamma\phi^{2}+(F\Gamma\phi^{2})^{\dagger}]
\nonumber\\
&&\rightarrow
\frac{g^{2}}{2!}[4\phi\phi\langle F\Gamma\phi\rangle
\langle\Gamma\phi F\rangle+4\phi^{\dagger}\phi^{\dagger}
\langle F^{\dagger}(\Gamma\phi)^{\dagger}\rangle
\langle(\Gamma\phi)^{\dagger}F^{\dagger}\rangle\nonumber\\
&&+8\phi\phi^{\dagger}\langle FF^{\dagger}\rangle
\langle\Gamma\phi(\Gamma\phi)^{\dagger}\rangle]\nonumber\\
&&=\frac{g^{2}}{2!}\int[4\phi\phi\frac{a^{2}m\Gamma^{2}}
{H^{2}+(am\Gamma)^{2}}\frac{a^{2}m\Gamma^{2}}
{H^{2}+(am\Gamma)^{2}}\nonumber\\
&&+4\phi^{\dagger}\phi^{\dagger}\frac{a^{2}m\Gamma^{2}}
{H^{2}+(am\Gamma)^{2}}\frac{a^{2}m\Gamma^{2}}
{H^{2}+(am\Gamma)^{2}}\nonumber\\
&&+8\phi\phi^{\dagger}\frac{-H^{2}}{H^{2}+(am\Gamma)^{2}}
\frac{a^{2}\Gamma^{2}}{H^{2}+(am\Gamma)^{2}}]\nonumber\\
&&=-4g^{2}a^{2}\phi\phi^{\dagger}\int\frac{\Gamma^{2}}{H^{2}
+(am\Gamma)^{2}}
\nonumber\\
&&+\frac{(mg)^{2}a^{4}}{2!}\int[8\phi\phi^{\dagger}
\frac{\Gamma^{2}}
{H^{2}+(am\Gamma)^{2}}\frac{\Gamma^{2}}
{H^{2}+(am\Gamma)^{2}}\nonumber\\
&&+4\phi^{\dagger}\phi^{\dagger}\frac{\Gamma^{2}}
{H^{2}+(am\Gamma)^{2}}\frac{\Gamma^{2}}
{H^{2}+(am\Gamma)^{2}}\nonumber\\
&&+4\phi\phi\frac{\Gamma^{2}}{H^{2}+(am\Gamma)^{2}}
\frac{\Gamma^{2}}{H^{2}+(am\Gamma)^{2}}].
\end{eqnarray}
The first term is quadratically divergent, and the remaining 
3 terms are logarithmically divergent.
These terms, when evaluated at vanishing external momenta, 
precisely cancel the fermion contributions (4.4).

\subsubsection{Triple Couplings}

The contributions from scalar
one-loop diagrams to scalar triple couplings are generated by
\begin{eqnarray}
\frac{g^{3}}{3!}\{[F\Gamma\phi^{2}
+(\phi^{\dagger})^{2}\Gamma F^{\dagger}]
[F\Gamma\phi^{2}
+(\phi^{\dagger})^{2}\Gamma F^{\dagger}]
[F\Gamma\phi^{2}
+(\phi^{\dagger})^{2}\Gamma F^{\dagger}]\}.
\end{eqnarray}
It is confirmed that the combinations $\phi\phi\phi$ and
$\phi^{\dagger}\phi^{\dagger}\phi^{\dagger}$ are convergent.
On the other hand, the combinations $\phi^{\dagger}\phi\phi$
and  $\phi^{\dagger}\phi^{\dagger}\phi$ are logarithmically
divergent. We thus set the external momenta at $0$ to 
extact the coefficients of divergent terms. For example,
we have
\begin{eqnarray}
&&\frac{g^{3}}{3!}[(\phi^{\dagger})^{2}\Gamma F^{\dagger}]
[F\Gamma\phi^{2}]
[F\Gamma\phi^{2}]\nonumber\\
&&\rightarrow
\frac{g^{3}}{3!}[16\phi^{\dagger}\phi\phi 
\langle F\Gamma\Gamma F^{\dagger}\rangle
\langle F\Gamma\phi\rangle
\langle\phi\phi^{\dagger}\rangle]\nonumber\\
&&=\frac{16g^{3}}{3!}\phi^{\dagger}\phi\phi
\int[\frac{-\Gamma^{2}H^{2}}{H^{2}+(ma\Gamma)^{2}}
\frac{-ma^{2}\Gamma^{2}}{H^{2}+(ma\Gamma)^{2}}
\frac{a^{2}}{H^{2}+(ma\Gamma)^{2}}]\\
&&=\frac{16g^{3}ma^{4}}{3!}\phi^{\dagger}\phi\phi
\int[\frac{H^{2}}{H^{2}+(ma\Gamma)^{2}}
\frac{\Gamma^{2}}{H^{2}+(ma\Gamma)^{2}}
\frac{\Gamma^{2}}{H^{2}+(ma\Gamma)^{2}}].\nonumber
\end{eqnarray}
Since we have 3 terms of this combination, we have the 
numerical coefficient
\begin{equation}
\frac{16g^{3}ma^{4}}{3!}\times 3=\frac{48g^{3}ma^{4}}{3!}
\end{equation}
which precisely cancels the fermion contribution (4.9), as 
required.

\subsubsection{Quartic Couplings}

The contributions from scalar
one-loop diagrams to quartic couplings are generated by
\begin{eqnarray}
&&\frac{g^{4}}{4!}[F\Gamma\phi^{2}
+(\phi^{\dagger})^{2}\Gamma F^{\dagger}]
[F\Gamma\phi^{2}
+(\phi^{\dagger})^{2}\Gamma F^{\dagger}]\nonumber\\
&&\times
[F\Gamma\phi^{2}
+(\phi^{\dagger})^{2}\Gamma F^{\dagger}][F\Gamma\phi^{2}
+(\phi^{\dagger})^{2}\Gamma F^{\dagger}]\}.
\end{eqnarray}
It is confirmed that only the combination 
$\phi^{\dagger}\phi^{\dagger}\phi\phi$
is  logarithmically divergent. We thus set the external momenta 
at $0$ to 
extact the coefficients of divergent terms.
We thus  have
\begin{eqnarray}
&&\frac{g^{4}}{4!}\frac{4\times3}{2!}\times2^{4}\times 2
\phi^{\dagger}\phi^{\dagger}\phi\phi
[ \langle F\Gamma\Gamma F^{\dagger}\rangle
\langle\phi\phi^{\dagger}\rangle
\langle F\Gamma\Gamma F^{\dagger}\rangle
\langle\phi\phi^{\dagger}\rangle]\nonumber\\
&&=\frac{g^{4}}{4!}\frac{4\times3}{2!}\times2^{4}\times 2
\phi^{\dagger}\phi^{\dagger}\phi\phi\nonumber\\
&&\times\int[\frac{-\Gamma^{2}H^{2}}{H^{2}+(ma\Gamma)^{2}}
\frac{a^{2}}{H^{2}+(ma\Gamma)^{2}}
\frac{-\Gamma^{2}H^{2}}{H^{2}+(ma\Gamma)^{2}}
\frac{a^{2}}{H^{2}+(ma\Gamma)^{2}}]\nonumber\\
&&=\frac{(ga)^{4}}{4!}\frac{4\times3}{2!}\times2^{4}\times 2
\phi^{\dagger}\phi^{\dagger}\phi\phi\\
&&\times\int[\frac{H^{2}}{H^{2}+(ma\Gamma)^{2}}
\frac{\Gamma^{2}}{H^{2}+(ma\Gamma)^{2}}
\frac{H^{2}}{H^{2}+(ma\Gamma)^{2}}
\frac{\Gamma^{2}}{H^{2}+(ma\Gamma)^{2}}]\nonumber
\end{eqnarray}
where the last numerical coefficient $2$ comes from the 
two possible ways to have 
$\langle F\Gamma\Gamma F^{\dagger}\rangle$.
The numerical coefficient
\begin{equation}
\frac{(ga)^{4}}{4!}\frac{4\times3}{2!}\times2^{4}\times 2
=\frac{(2ga)^{4}}{2!}
\end{equation}
is precisely what we need to cancel the fermion contribution 
(4.12).

It is confirmed that all the higher order couplings are 
finite.

\subsection{Fermion Self-energy Correction}

The self-energy correction to the fermion for general external 
momentum is given by
\begin{eqnarray}
&&\frac{1}{2!}[2g\bar{\psi}(P_{+}\phi\hat{P}_{+}
+P_{-}\phi^{\dagger}\hat{P}_{-})\psi]
[2g\bar{\psi}(P_{+}\phi\hat{P}_{+}
+P_{-}\phi^{\dagger}\hat{P}_{-})\psi]\nonumber\\
&&\rightarrow
\frac{2(2g)^{2}}{2!}\bar{\psi}P_{+}[\int \hat{P}_{+}\frac{-1}
{D+m\gamma_{5}\Gamma_{5}}
P_{-}\frac{a^{2}}{H^{2}+(am\Gamma)^{2}}]\hat{P}_{-}\psi\nonumber\\
&&+\frac{2(2g)^{2}}{2!}\bar{\psi}P_{-}[\int \hat{P}_{-}
\frac{-1}{D+m\gamma_{5}\Gamma_{5}}P_{+}
\frac{a^{2}}{H^{2}+(am\Gamma)^{2}}]\hat{P}_{+}\psi
\nonumber\\
&&=-4g^{2}a^{3}\bar{\psi}P_{+}[\int \Gamma_{5}
\frac{H+am\Gamma_{5}}
{H^{2}+(am\Gamma_{5})^{2}}
\frac{1}{H^{2}+(am\Gamma)^{2}}]P_{-}\Gamma_{5}\psi\nonumber\\
&&+4g^{2}a^{3}\bar{\psi}P_{-}[\int \Gamma_{5}
\frac{H+am\Gamma_{5}}
{H^{2}+(am\Gamma_{5})^{2}}
\frac{1}{H^{2}+(am\Gamma)^{2}}]P_{+}\Gamma_{5}\psi\nonumber\\
&&=-4g^{2}a^{3}\bar{\psi}P_{+}[\int \frac{\Gamma_{5}H}
{H^{2}+(am\Gamma_{5})^{2}}
\frac{1}{H^{2}+(am\Gamma)^{2}}]P_{-}\Gamma_{5}\psi\nonumber\\
&&+4g^{2}a^{3}\bar{\psi}P_{-}[\int \frac{\Gamma_{5}H}
{H^{2}+(am\Gamma_{5})^{2}}
\frac{1}{H^{2}+(am\Gamma)^{2}}]P_{+}\Gamma_{5}\psi
\end{eqnarray}
where we used the fact that $H^{2}$ and $\Gamma_{5}^{2}$ are
 independent of
$\gamma$ matrices and $P_{\pm}\hat{P}_{\pm}
=\pm P_{\pm}\Gamma_{5}$ for the 
external lines. 
  
By noting the relations
\begin{equation}
P_{\pm}\Gamma_{5}H P_{\mp}=P_{\pm}\gamma_{5}H P_{\mp}
=\pm P_{\pm}H P_{\mp}
\end{equation}
the above amplitude is written as 
\begin{eqnarray}
&&-4g^{2}a^{3}\bar{\psi}P_{+}[\int \frac{H}
{H^{2}+(am\Gamma_{5})^{2}}
\frac{1}{H^{2}+(am\Gamma)^{2}}]P_{-}\Gamma_{5}\psi\nonumber\\
&&-4g^{2}a^{3}\bar{\psi}P_{-}[\int \frac{H}
{H^{2}+(am\Gamma_{5})^{2}}
\frac{1}{H^{2}+(am\Gamma)^{2}}]P_{+}\Gamma_{5}\psi.
\end{eqnarray}
Because of the projection operators $P_{\pm}$, which sandwich
the amplitude with integral over the loop momentum, only 
the amplitude linear in $\gamma^{\mu}$ matrices survives the 
projection. We thus have no terms corresponding to the mass 
term. The mass renormalization vanishes when renormalized at 
vanishing external momentum.

\subsection{Correction to Fermionic Vertex}

The one-loop correction to the fermionic vertex is finite
as is seen in , for example,
\begin{eqnarray}
&&(2g)^{3}a^{4}\bar{\psi}P_{+}[\int\hat{P}_{+}
\frac{H+ma\Gamma_{5}}{H^{2}+(ma\Gamma_{5})^{2}}\gamma_{5}
P_{+}\phi\hat{P}_{+}\frac{H+ma\Gamma_{5}}
{H^{2}+(ma\Gamma_{5})^{2}}\gamma_{5}P_{-}
\frac{1}{H^{2}+(ma\Gamma)^{2}}]\hat{P}_{-}\psi\nonumber\\
&&=(2g)^{3}a^{4}\bar{\psi}P_{+}[\int\Gamma_{5}
\frac{H+ma\Gamma_{5}}{H^{2}+(ma\Gamma_{5})^{2}}\gamma_{5}
P_{+}\phi\Gamma_{5}\frac{H+ma\Gamma_{5}}
{H^{2}+(ma\Gamma_{5})^{2}}\gamma_{5}P_{-}
\frac{1}{H^{2}+(ma\Gamma)^{2}}]\hat{P}_{-}\psi\nonumber\\
&&=-(2g)^{3}ma^{5}\bar{\psi}P_{+}[\int
\frac{\Gamma^{2}_{5}}{H^{2}+(ma\Gamma_{5})^{2}}
P_{+}\phi\frac{\Gamma_{5}H}
{H^{2}+(ma\Gamma_{5})^{2}}P_{-}
\frac{1}{H^{2}+(ma\Gamma)^{2}}]\hat{P}_{-}\psi\nonumber\\
&&=-(2g)^{3}ma^{5}\bar{\psi}P_{+}[\int
\frac{\Gamma^{2}_{5}}{H^{2}+(ma\Gamma_{5})^{2}}
P_{+}\phi\frac{H}
{H^{2}+(ma\Gamma_{5})^{2}}P_{-}
\frac{1}{H^{2}+(ma\Gamma)^{2}}]\hat{P}_{-}\psi
\end{eqnarray}
which is convergent and vanishes at vanishing external
momenta after a symmetric integral over the internal momentum.  

Our analysis so far establishes the non-renormalization (i.e.,
the absence of even finite renormalization) of 
the mass and interaction terms to the one-loop order in 
perturbation theory for a finite lattice spacing $a$ when 
renormalized at {\em vanishing momenta}.  

\subsection{Renormalization of Kinetic Terms}

The one-loop correction to the ``kinetic term'' $FF^{\dagger}$
is given by
\begin{eqnarray}
&&\frac{g^{2}}{2!}[F\Gamma(\phi)^{2}
+(F\Gamma(\phi)^{2})^{\dagger}]^{2}\nonumber\\
&&\rightarrow
\frac{g^{2}}{2!}[4F\Gamma\langle\phi\phi^{\dagger}\rangle
\langle\phi\phi^{\dagger}\rangle\Gamma^{\dagger}F^{\dagger}]
\nonumber\\
&&=2g^{2}a^{4}F\Gamma[\int 
\frac{1}{H^{2}+(am\Gamma)^{2}}
\frac{1}{H^{2}+(am\Gamma)^{2}}]
\Gamma F^{\dagger}
\end{eqnarray}
which is logarithmically divergent.

The  correction to the kinetic term of scalar particles
is given in (4.3)
\begin{equation}
g^{2}a^{2}Tr\phi^{\dagger}
\frac{H\Gamma_{5}}{H^{2}+(ma\Gamma_{5})^{2}}
\phi\frac{\Gamma_{5}H}{H^{2}+(ma\Gamma_{5})^{2}}
\end{equation}
where the quadratic divergence evaluated at vanishing external 
momentum is cancelled by the scalar loop diagrams. 

The fermion self-energy correction is given by (4.24)
\begin{eqnarray}
&&-4g^{2}a^{3}\bar{\psi}P_{+}[\int \frac{H}
{H^{2}+(am\Gamma_{5})^{2}}
\frac{1}{H^{2}+(am\Gamma)^{2}}]P_{-}\Gamma_{5}\psi\nonumber\\
&&-4g^{2}a^{3}\bar{\psi}P_{-}[\int \frac{H}
{H^{2}+(am\Gamma_{5})^{2}}
\frac{1}{H^{2}+(am\Gamma)^{2}}]P_{+}\Gamma_{5}\psi.
\end{eqnarray}
We have to examine if a universal wave function renormalization 
is sufficient
to remove the divergence from these 3 contributions.

One can in fact show that a uniform subtraction of logarithmic 
infinity
renders all these expressions finite.
The integral for $FF^{\dagger}$ (4.26) is written in more 
detail as
\begin{equation}
2g^{2}\Gamma(ap)[\int_{-\pi}^{\pi}\frac{d^{4}k}{(2\pi)^{4}} 
\frac{1}{H^{2}(k+ap)+(am\Gamma)^{2}(k+ap)}
\frac{1}{H^{2}(k)+(am\Gamma)^{2}(k)}]
\Gamma(ap).
\end{equation} 
where we chose the basic Brillouin zone at $\frac{-\pi}{a}<
k_{\mu}\leq \frac{\pi}{a}$ and rescaled the integration 
variables
as $a k_{\mu}\rightarrow k_{\mu}$.
If one renormalizes this expression at $p_{\mu}=0$, the wave 
function renormalization factor is defined by
\begin{equation}
Z=1-2g^{2}\int_{-\pi}^{\pi}\frac{d^{4}k}{(2\pi)^{4}} 
\frac{1}{H^{2}(k)+(am\Gamma)^{2}(k)}
\frac{1}{H^{2}(k)+(am\Gamma)^{2}(k)}
\end{equation}
by noting $\Gamma(0)=1$. This integral in $Z$ is logarithmically
divergent for $a\rightarrow 0$: The divergence appears as 
an infra-red divergence. The divergent part in (4.26) at 
vanishing external momentum is extracted by
considering  small $\delta, \ |\delta|\ll 1$, as 
\begin{eqnarray}
&&2g^{2}\int_{-\delta}^{\delta}\frac{d^{4}k}{(2\pi)^{4}} 
\frac{1}{H^{2}(k)+(am\Gamma)^{2}(k)}
\frac{1}{H^{2}(k)+(am\Gamma)^{2}(k)}
\nonumber\\
&&\simeq2g^{2}\int_{-\delta}^{\delta}\frac{d^{4}k}{(2\pi)^{4}} 
\frac{1}{k^{2}+(am)^{2}}
\frac{1}{k^{2}+(am)^{2}}\nonumber\\
&&=2g^{2}\int_{-\delta/a}^{\delta/a}\frac{d^{4}k}{(2\pi)^{4}} 
\frac{1}{k^{2}+m^{2}}
\frac{1}{k^{2}+m^{2}}.
\end{eqnarray}
Note that $H^{2}\sim O(1)$ for the domain of species doublers
and thus no infrared divergence.

The divergent part of the self-energy of scalar particles (4.27)
is extracted by considering small $p_{\mu}$ and 
$\delta$ as
\begin{eqnarray}
&&\frac{g^{2}}{a^{2}}Tr\int^{\delta}_{-\delta}
\frac{d^{4}k}{(2\pi)^{4}}
\frac{H(k+ap)\Gamma_{5}(k+ap)}
{H^{2}(k+ap)+(ma\Gamma_{5})^{2}(k+ap)}\times
\frac{\Gamma_{5}(k)H(k)}{H^{2}(k)+(ma\Gamma_{5})^{2}(k)}
\nonumber\\
&&\simeq\frac{g^{2}}{a^{2}}Tr\int^{\delta}_{-\delta}
\frac{d^{4}k}{(2\pi)^{4}}
\frac{-\gamma_{5}(\kslash+a\pslash)\gamma_{5}}
{(k+ap)^{2}+(ma)^{2}}\times
\frac{-\kslash}{k^{2}+(ma)^{2}}\nonumber\\
&&=4\frac{g^{2}}{a^{2}}\int^{\delta}_{-\delta}
\frac{d^{4}k}{(2\pi)^{4}}
\frac{k^{2}+akp}
{k^{2}+2akp+(ap)^{2}+(ma)^{2}}\times
\frac{1}{k^{2}+(ma)^{2}}\nonumber\\
&&\rightarrow
4\frac{g^{2}}{a^{2}}\int^{\delta}_{-\delta}
\frac{d^{4}k}{(2\pi)^{4}}[-\frac{(k^{2}+akp)(2akp+(ap)^{2})}
{(k^{2}+(ma)^{2})^{3}}\nonumber\\
&&+\frac{(k^{2}+akp)(2akp+(ap)^{2})^{2}}{(k^{2}+(ma)^{2})^{4}}]
\nonumber\\
&&=4\frac{g^{2}}{a^{2}}\int^{\delta}_{-\delta}
\frac{d^{4}k}{(2\pi)^{4}}[-\frac{k^{2}(ap)^{2}+2(akp)^{2}}
{(k^{2}+(ma)^{2})^{3}}
+\frac{4k^{2}(akp)^{2}}{(k^{2}+(ma)^{2})^{4}}]
\nonumber\\
&&=4g^{2}p^{2}\int^{\delta}_{-\delta}
\frac{d^{4}k}{(2\pi)^{4}}[-\frac{3}{2}\frac{k^{2}}
{(k^{2}+(ma)^{2})^{3}}
+\frac{k^{4}}{(k^{2}+(ma)^{2})^{4}}]
\nonumber\\
&&\rightarrow
2g^{2}\int^{\delta}_{-\delta}
\frac{d^{4}k}{(2\pi)^{4}}\frac{1}
{(k^{2}+(ma)^{2})^{2}}\times (-p^{2})
\end{eqnarray}
where we used the fact that the $p$-independent terms
are cancelled  by the mass correction arising from the scalar
particle one-loop diagrams. 
We also took account of a symmetric 
integral over
$k_{\mu}$.
Note that our $\gamma^{\mu}$ 
matrices are anti-hermitian, and  $k^{2}\geq 0$.
 We thus have the same divergent
part as that of the $FF^{\dagger}$ term in (4.31).

Similarly, the divergent part of the fermion self-energy
correction (4.28) is extracted by considering small $p_{\mu}$ 
and $\delta$.
\begin{eqnarray}
&&-4\frac{g^{2}}{a}P_{+}[\int_{-\delta}^{\delta}
\frac{d^{4}k}{(2\pi)^{4}} \frac{H(k+ap)}
{H^{2}(k+ap)+(am\Gamma_{5})^{2}(k+ap)}\times
\frac{1}{H^{2}(k)+(am\Gamma)^{2}(k)}]\nonumber\\
&&\times P_{-}\Gamma_{5}(ap)
\nonumber\\
&&-4\frac{g^{2}}{a}P_{-}[\int_{-\delta}^{\delta}
\frac{d^{4}k}{(2\pi)^{4}} \frac{H(k+ap)}
{H^{2}(k+ap)+(am\Gamma_{5})^{2}(k+ap)}\times
\frac{1}{H^{2}(k)+(am\Gamma)^{2}(k)}]\nonumber\\
&&\times P_{+}\Gamma_{5}(ap)
\nonumber\\
&&\simeq
-4\frac{g^{2}}{a}\int_{-\delta}^{\delta}
\frac{d^{4}k}{(2\pi)^{4}}
\frac{-\gamma_{5}
(\kslash+a\pslash)}
{(k+ap)^{2}+(am)^{2}}\times
\frac{1}{k^{2}+(am)^{2}}\gamma_{5}
\nonumber\\
&&\rightarrow
-4\frac{g^{2}}{a}\int_{-\delta}^{\delta}
\frac{d^{4}k}{(2\pi)^{4}}
[\frac{a\pslash}{(k^{2}+(am)^{2})^{2}}
-\frac{1}{2}\frac{a\pslash k^{2}}{(k^{2}+(am)^{2})^{3}}]
\nonumber\\
&&\rightarrow
2g^{2}\int_{-\delta}^{\delta}\frac{d^{4}k}{(2\pi)^{4}}
\frac{1}{(k^{2}+(am)^{2})^{2}}\times(-\pslash)
\end{eqnarray}
by taking into account a symmetric integral over $k_{\mu}$.
We thus obtain the same divergent part as the other two 
wave function renormalization factors (4.31) and (4.32).

The finite part in wave function renormalization 
is, however, generally different for these three quantities,
which will cause a deviation from supersymmetry, such as 
the quadratic divergence of the scalar mass correction, in the 
two-loop order in perturbation theory.

\section{Discussion and Conclusion}

We have studied the implications of lattice chiral symmetry
satisfied by the Ginsparg-Wilson operators when applied to a 
regularization
of the supersymmetric Wess-Zumino model. 
We found a conflict between the lattice chiral symmetry
and the Euclidean Majorana condition for Yukawa couplings. 
(In the Weyl representation, we recognized a conflict 
between the lattice chiral symmetry and Yukawa couplings.)
We thus have basically two alternative choices: the first one is 
to impose lattice chiral symmetry but sacrifice the Euclidean 
Majorana condition, and the second one is to sacrifice lattice 
chiral symmetry but preserve the Euclidean Majorana condition. 
The latter approach is similar to the analysis of Bartels and 
Kramer\cite{bartels} on the basis of the conventional Wilson 
fermion, where one-loop level non-renormalization of the 
superpotential
 has been shown: We have also confirmed this property for the 
Lagrangian (2.28). In this paper, we examined in detail the 
case 
(2.17) where the lattice chiral symmetry together with naive 
Bose-Fermi symmetry is imposed. This scheme incorporates
a SUSY-like symmetry in the free part of the Lagrangian and 
preserves one-loop level non-renormalization of the 
superpotential for finite lattice spacing, independently of the 
parameter $k$ in (1.2) and independently of the detailed 
construction of lattice Dirac operators. Both schemes however 
fail in 
preserving the uniform wave function renormalization for all 
the field variables. This generally causes the failure in the 
cancellation of quadratic divergences in the two-loop
order in perturbation theory. 
These two schemes are thus on a similar footing.

From a view point of practical diagramatical calculations,
our analysis illustrates a transparent treatment of 
various diagrams on the basis of algebraic properties 
without recourse to the detailed construction of lattice
Dirac operators. The actual calculations are almost identical 
to those in continuum theory, and the 
power counting of various daigrams agrees with that of the 
continuum theory, as was shown in Section 3; this agreement 
of the power counting does not hold for the Lagrangian (2.28).
Although our analysis is perturbative, a perturbative lattice 
analysis of Yukawa couplings
is expected to be important when one applies a non-perturbative 
analysis to the gauge theoretical sector of a supersymmetric 
model of elementary particles, just to preserve the consistency 
of various sectors of a single model. When one includes a
background gauge field into our treatment, one need to 
 covariantize the operator $D^{\dagger}D$ in the bosonic
sector by a lattice analogue of
\begin{equation}
\partial_{\mu}\rightarrow \partial_{\mu}-igA_{\mu}  
\end{equation}
independently\footnote{Note that $D^{\dagger}D$ after 
covariantizing $D$ and a direct covariantization of 
$D^{\dagger}D$ are generally different. We also note that 
the locality of $D$ for $k>0$ with dynamical gauge field has 
not been established yet\cite{fujikawa3}.} of the fermionic 
sector defined by $D$, to
avoid the appearance of the Pauli term, for example, in 
$D^{\dagger}D$.

There exists an argument for non-renormalization theorem on the 
basis of the stability of supersymmetry and holomorphic 
properties\cite{seiberg}. The holomorphic
properties are partly related to chiral symmetry and partly
related to a clear separation between the kinetic (K\"{a}hler)
and superpotential terms. Our analysis indicates that 
these two requirements are in conflict in the lattice 
regularization known so far\footnote{For the Lagrangian (2.28), 
the term ${{{\cal L}_{M}}_{(2)}}_{pot}$ (2.32) breaks $U(1)$ in 
$U(1)\times U(1)_{R}$ symmetry even 
for $m=g=0$ and thus the argument in \cite{seiberg} does not 
work.}. As for the  supersymmetry
algebra, it is known that the Leibniz rule generally 
fails in the lattice difference operation\cite{dondi}. 
Apparently, much need to be learned before we obtain a 
coherent picture of the lattice regularization of
supersymmetry.\\    

We thank H. So and H. Suzuki for helpful comments.

\end{document}